\documentclass[twocolumn]{revtex4-1}

\usepackage{amsmath,amssymb,graphicx} 
\usepackage{subfigure}
\usepackage{amsfonts}
\usepackage[english]{babel}
\usepackage[applemac]{inputenc}
\usepackage{color}
\usepackage{amsthm}
\usepackage{epsfig}
\usepackage{subfigure}
\usepackage{cancel}
\usepackage{verbatim}
\usepackage{eepic}

\begin{document}
\title{Antinematic local order in dendrimer liquids}
%\shorttitle{Antinematic local order in dendrimer liquids} %Insert here a short version of the title if it exceeds 70 characters

\author{Ioannis A. Georgiou}
\affiliation{Institute for Theoretical Physics, Technische Universit\"at Wien, Wiedner Hauptstra{\ss}e 8-10, A-1040 Vienna, Austria}

\author{Primo{\v z} Ziherl}
\affiliation{Faculty of Mathematics and Physics, University of Ljubljana, Jadranska 19, SI-1000 Ljubljana, Slovenia}
\affiliation{Jo\v zef Stefan Institute, Jamova 39, SI-1000 Ljubljana, Slovenia}
	
\author{Gerhard Kahl}
\affiliation{Institute for Theoretical Physics and Center for Computational Materials Science, Technische Universit\" at Wien, Wiedner Hauptstra{\ss}e 8-10, A-1040 Vienna, Austria}

\begin{abstract}
We use monomer-resolved numerical simulations to study the positional and orientational structure of a dense dendrimer solution, focusing on the effects prolate shape and deformability of the dendrimers on the short-range order. Our results provide unambiguous evidence that the nearest-neighbor shell of a tagged particle consists of a mixture of crossed, side-by-side, side-to-end, and end-to-end pair configurations, imposing {\it antinematic} rather than nematic order observed in undeformable rodlike particles. This packing pattern persists even at densities where particle overlap becomes sizable. We demonstrate that the antinematic arrangement is compatible with the A15 crystal lattice reported in several dendrimer compounds.\newline\newline
\newline
\newline
\end{abstract}

%\pacs{47.57.J-}{Colloids; complex fluids}
%\pacs{61.20.-p}{Structure of liquids}
%\pacs{61.20.Ja}{Computer modeling and simulation; liquid structure}

%************************************************************************************************************
\newcommand{\LL}{\mbox{\Thicklines \,\line(0,1){7}\line(1,0){7}}\,}
\newcommand{\TT}{\mbox{\Thicklines \,\line(1,0){7}\hspace*{-3.5pt}\line(0,1){7}\hspace*{3.5pt}}\,}
\newcommand{\XX}{\mbox{\Thicklines \,\raisebox{3.5pt}{\line(1,0){7}}\!\!\line(0,1){7}\hspace*{3.5pt}}\,}
\newcommand{\II}{\mbox{\Thicklines \,\line(0,1){7}\hspace*{3.5pt}\line(0,1){7}}\,}
\newcommand{\PP}{\mbox{\Thicklines \,\raisebox{3.5pt}{\line(1,0){7}\hspace*{3.5pt}\line(1,0){7}}}\,}
\newcommand{\Idot}{\mbox{\Thicklines \,\line(0,1){7}\,\,\raisebox{5.5pt}{\LARGE .}}\,}
\newcommand{\antipara}{\mbox{\Thicklines \,\line(0,1){7}\, \raisebox{3.5pt}{\line(1,0){7}}\,\raisebox{3pt}{\LARGE .}\,\line(0,1){7}\,\raisebox{3pt}{\LARGE .}\,\line(0,1){7}\, \raisebox{3.5pt}{\line(1,0){7}}}\,}
\newcommand{\DD}{\mbox{\Thicklines \,\raisebox{3.5pt}{\line(1,0){7}}\hspace*{-3.5pt}\raisebox{0pt}{\line(1,0){7}}}\,}
%\newcommand{\Idot}{\mbox{\Thicklines \,\line(0,1){7}\,\,\raisebox{6pt}{.}}\,}
%************************************************************************************************************

\maketitle

% INTRODUCTION

\section{Introduction}

Dendrimers are branched macromolecules with a tree-like structure. This particular architecture is the result of a controlled, step-by-step synthesis where -- starting from a pair of bonded tri-functional monomers -- the macromolecule grows by adding core monomers generation by generation. The outermost, so-called shell monomers can be decorated by suitable end groups. The first report on the synthesis of dendrimers by V\"ogtle and co-workers in 1978~\cite{Buhleier78} received little attention but the interest in these materials increased considerably after their potential applications were pointed out~\cite{Tomalia85}. 

% DENDRIMERS HALFWAY BETWEEN LINEAR POLYMERS AND COLLOIDS

With a compact convex shape determined by their architecture, dendrimers are in many ways more similar to colloidal particles than to spread-out, random-coil linear polymers -- yet they are penetrable like random coils such that two dendrimers can overlap.  Much like colloidal particles, dendrimers readily crystallize but many lattices observed (e.g., A15 and $\sigma$ phases~\cite{Balagurusamy97,Zeng04}) are untypical for classical colloids which usually form face- and body-centered cubic crystals. Some aspects of this unique behavior may be related to the dendrimer shape: The early-generation dendrimers can be viewed as prolate ellipsoids, the molecular elongation decreasing with the generation number (cf. Fig.~5 of Ref.~\cite{Maiti04}). The aspect ratio of dendrimers is smaller than in linear polymers~\cite{Solc71} but still quite large and may exceed four in the first, innermost generation~\cite{Naylor89}. 

It is natural to expect that the optimal packing mode of dendrimers will depend on their shape and deformability. Indeed, atomistic simulations revealed that at large densities considerable interpenetration does take place~\cite{Li04} leading to the A15 cubic lattice as seen experimentally. Complementary to this prediction are theoretical studies of penetrable ellipsoids interacting with anisotropic Gaussian repulsion. If forced into alignment, they form elongated lattices obtained, e.g., by stretching the body-centered cubic crystal along the [001], [110], or [111] directions~\cite{Prestipino07,Nikoubashman09}. This implies that parallel alignment of dendrimers is incompatible with the cubic symmetry and that the pattern of their relative orientation in the A15 lattice must be more complex.  

% FOURTH PARAGRAPH: "In this Letter,\ldots"

The existing body of experimental and theoretical results clearly shows that there exists a link connecting dendrimer shape and deformability with the open lattices such as the A15 and $\sigma$ phases. However, the workings of this link as well as its possible consequences beyond the stability of open crystal lattices remain poorly understood. To shed light on the complex interplay of dendrimer deformation and reorientation, it is worthwhile to develop a coarse-grained description of dendrimers where they are regarded as soft, anisometric particles. 

Here we use monomer-resolved numerical simulations to investigate the short-range structure of a dendrimer liquid and we interpret the results in terms of dendrimer shape, anisotropic positional order, and orientational order. We find that dendrimers align such that the long axes of most nearest neighbors are perpendicular, e.g., \antipara \ (where dots represent rods pointing into or out of the paper). This so-called antinematic local packing pattern~\cite{Sokalski84} is robust and reveals a new and deeper insight into the structure of dendrimer crystals. The possibility of antinematic order raises several interesting questions including the potential existence of antinematic liquid phase with long-range orientational order (qualitatively reminiscent of the cubatic phase hypothesized in rod-like polyelectrolytes bound by flexible cross-linking bonds~\cite{Borukhov05}) and the phase diagram of a dendrimer solution. A comprehensive analysis of these issues in terms of monomer-resolved models, which entails the exploration of a rather large parameter space, is beyond the scope of this exploratory study and has been relegated to future work.

% MODEL AND NUMERICAL METHODOLOGY

\section{The model}

To provide a coarse-grained yet accurate picture of interacting dendrimers, we resort to monomer-resolved simulations, sacrificing the atomistic details so as to maximize the number of dendrimers in the system and to minimize the finite-size effects. The model dendrimers studied here have already been explored in some detail~\cite{Mladek08,Lenz09}, and they are based on tri-functional monomers. The total number of monomers in a dendrimer of generation number $G$ reads $2(2^{G+1}-1)$, which includes $2^{G+1}-2$ core monomers and $2^{G+1}$ terminal shell monomers. Both core and shell monomers interact with the Morse potential and the bonds between them are represented by the finitely extensible nonlinear elastic (FENE) potential [Eqs.~(2) and (1) of Ref.~\cite{Mladek08}, respectively]. The core monomers are different from the shell monomers, the main differences being the deeper attractive well and the shorter bond length. The values of model parameters are listed in Table I using the notation of Ref.~\cite{Mladek08}; they are virtually identical to those used as the past studies so as to facilitate the comparison with other aspects of dendrimer behavior discussed there~\cite{Mladek08,Lenz09}. Unless indicated otherwise all results reported here pertain to the fourth generation ($G=4$) dendrimers. 
\begin{table}[h]
  \begin{tabular}{cccccccc}
   \multicolumn{1}{c}{} & \multicolumn{3}{c}{Morse} & $\!$ & \multicolumn{3}{c}{FENE} \\ \cline{2-4} \cline{6-8} 
   \multicolumn{1}{c}{} & $\!\epsilon/k_{\rm B}T\!$ & $\! ad_{\rm CC}\!$ & $\! d/d_{\rm CC}\!\!$ & $\!\!$ & $\!\! Kd^2_{\rm CC}\!$ & $\! l^0/d_{\rm CC}\!$ & $\! R/d_{\rm CC}\phantom{\Big|}\!\!\!$	 \\ \hline
  CC   & 0.714 & 6.4  & 1.00 & $\!\!$ & 40 &  1.8750 & 0.3750\\
  CS   & 0.014 & 19.2 & 1.25 & $\!\!$ & 20 &  2.8125 & 0.5625\\
  SS   & 0.014 & 19.2 & 1.50
 \end{tabular}
  
\caption{Parameters of the core-core (CC), core-shell (CS), and shell-shell (SS) inter-monomer Morse potential ($\epsilon$, $a$, and $d$) and of the core-core and core-shell FENE bonds ($K$, $l^0$, and $R$) [see Eqs.~(2) and (1) of Ref.~\cite{Mladek08}] in our dendrimers. Lengths are given in units of $d_{\rm CC}$ whereas $\epsilon$ is in units of $k_{\rm B}T$. For all Morse interactions we truncate and shift the potential at a cut-off radius $r_c=2.8d_\text{CC}$.}
\label{tab:potparams}
\end{table}

Using standard $NVT$ Monte Carlo simulations, we obtain the equilibrium structure of a single dendrimer, a dendrimer pair, and an ensemble of $N=220$ dendrimers in a cubic box. Our choice of $N$ is a compromise between accuracy and computational effort: (i)~data obtained in smaller ensembles are essentially identical so that the short-range structure of the $N=220$ system is representative of a bulk liquid; (ii)~to ensure a sufficient accuracy (e.g., for probability distribution functions) we had to perform for each state point many independent runs extending over rather long time intervals, which was feasible at this value of $N$.

Starting from several independent initial configurations at high temperatures, we cool the system using a simulated annealing protocol to reach the desired temperature $T$ such that $k_{\rm B}T$ is $1.4\epsilon_{\rm CC}$, $\epsilon_{\rm CC}$ being the depth of the core-core attractive potential. The protocol employed depends on density, which is encoded by the packing fraction 
\begin{equation}
\phi_m=\frac{V_m}{V}
\end{equation} 
defined as the bare volume of monomers $V_m$~\cite{remark3} divided by the total volume $V$. The number of different realizations analyzed ranges from 1000 at the largest packing fraction considered $\phi_m=0.248$
to 500 at $\phi_m=0.199$ and 20 at $\phi_m=0.095$. The positions of monomers are recorded for at least $5\times 10^6$ Monte-Carlo sweeps. 

% SHAPE OF ISOLATED DENDRIMERS

\section{Shape of an isolated dendrimer}

We quantify the shape of dendrimers by computing the radius of gyration tensor ${\cal S}_{ik}=\left\langle x_ix_k\right\rangle,$ where $x_i$ is the $i$-th coordinate of a monomer in the dendrimer's center-of-mass system, and the average is over all monomers in a dendrimer and over $2\times10^4$ and $2\times10^5$ frames for a single dendrimer and for a pair, respectively. From the eigenvalues of ${\cal S}$, denoted by $E_1,E_2,$ and $E_3$ and arranged in descending order, we compute the asphericity ($b$) and the acylindricity ($c$) 
\begin{equation}
b = \frac{E_1-\left(E_2+ E_3\right)/2}{R_g^2}
%\end{equation}
%\begin{equation}
~~~~~~~~
c = \frac{E_2 - E_3}{R_g^2}
\end{equation}
as well as the radius of gyration defined by $R_g^2=E_1+E_2+E_3$~\cite{Solc71,Theodorou85}. We note that 
$b$ and $c$ are not the only possible choice of shape measures~\cite{Rudnick86}.

In spherical dendrimers $E_1=E_2=E_3$ so that $b$ and $c$ vanish whereas in non-spherical ones they do not. After analyzing dendrimers of generations two to ten, we find that they can be thought of a non-axisymmetric prolate ellipsoids which become increasingly more spherical as $G$ is increased. In particular, in {\it isolated} ($G=2$)-dendrimers $b=0.35 \pm 0.12$ and $c=0.15 \pm 0.07$ (which corresponds to semiaxes ratio of 2.03:1.44:1) whereas the ($G=10$)-dendrimers are almost spherical with  $b=0.061\pm0.012$ and $c=0.022\pm0.014$~\cite{remark1}. In the case of $(G=4)$-dendrimers the semiaxes ratio is 1.68:1.28:1.

% SHAPE OF TWO INTERACTING DENDRIMERS

\section{Two interacting dendrimers}

\begin{figure}[t]
\begin{center}
\includegraphics[width=0.9\columnwidth]{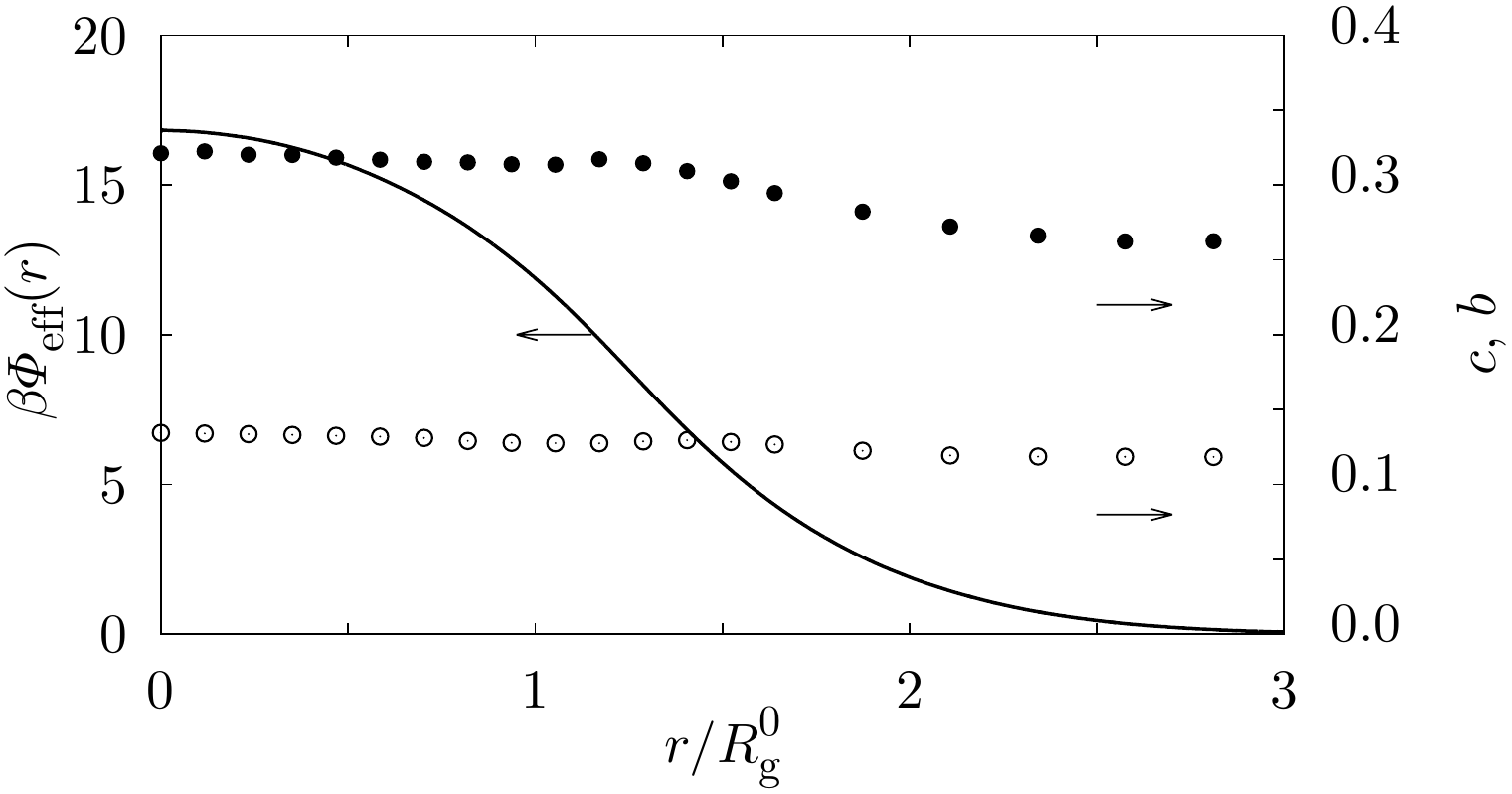}
\caption{Effective potential, $\beta \Phi_{\rm eff}(r)$, of a pair of ($G=4$)-dendrimers (solid line), and their asphericity, $b$, and acylindricity, $c$ (filled and open circles, respectively, both shown on the secondary vertical axis) vs. center-to-center separation.}
\label{Phibc} 
\end{center}
\end{figure}

The effective potential of dendrimers and their shape must be strongly interrelated. The solid line in Fig.~\ref{Phibc} shows the effective interaction energy, $\beta \Phi_{\rm eff}(r)$, of two ($G=4$)-dendrimers in the zero-density limit computed using umbrella sampling~\cite{Frenkel02}. The pair potential can be approximated by the generalized exponential model of index $\approx2.6$~\cite{Mladek06} suggesting that the dendrimers may form clusters of overlapping particles~\cite{Likos01}. Also plotted in Fig.~\ref{Phibc} are the dendrimer asphericity and acylindricity as functions of the center-to-center distance. We observe a sizeable increase of asphericity $b$ of $\approx 22 \%$ as $r/R^0_g$ is decreased from 2.5 to 1.5, $R_g^0$ being the radius of gyration of an isolated dendrimer. On the other hand, the acylindricity $c$ essentially does not deviate from its value in isolated particles. Concomitantly, the effective interaction increases by $\approx 30 \%$ of the potential at complete overlap $\beta\Phi_{\rm eff}(r=0)$. These results imply both in terms of shape and in terms of energy that a partial interpenetration takes place for $1.5\lesssim r/R^0_g \lesssim 2.5$ and that for $r/R_g^0\lesssim 1$ one can speak of the complete overlap regime where neither shape nor pair interaction depend very much on the center-to-center separation.

\begin{figure*}[ht]
\begin{center}
\includegraphics[width=1.9\columnwidth]{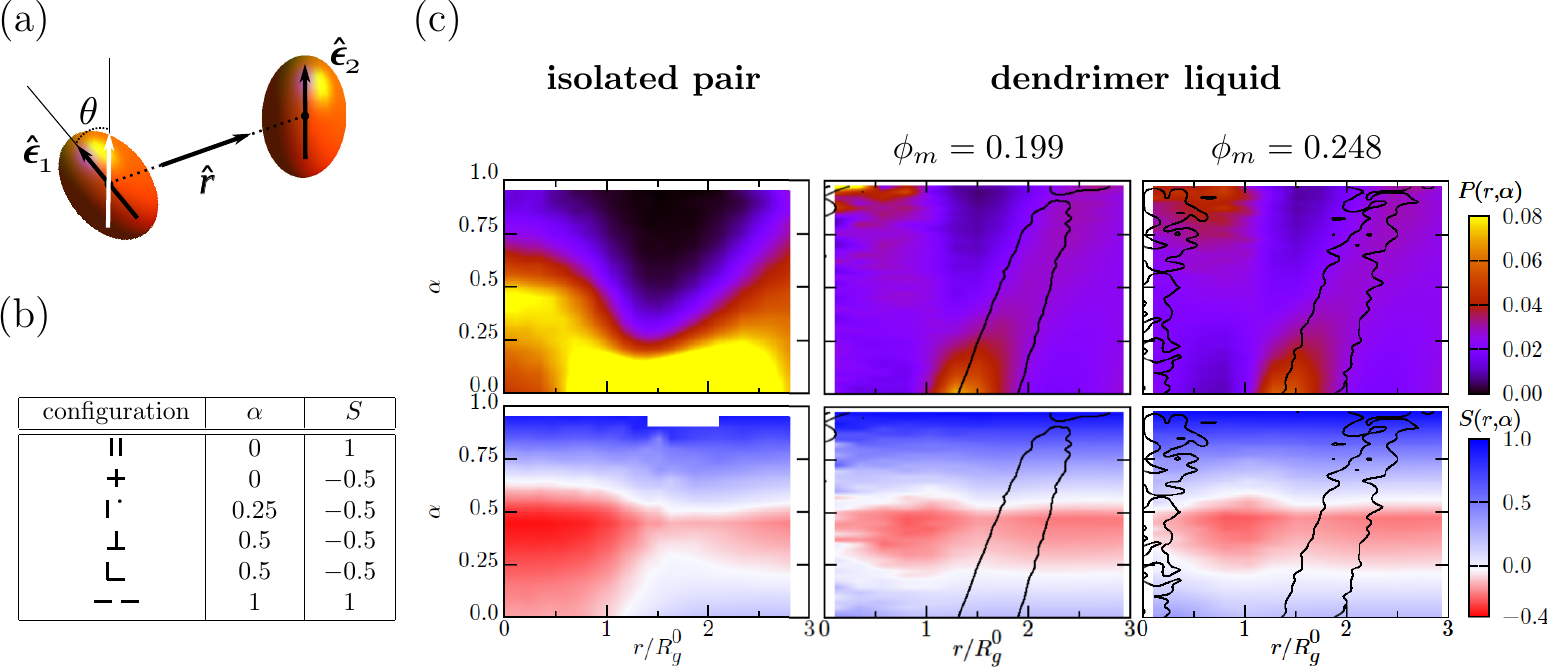}
\caption{Positional and orientational order of two interacting dendrimers. Panel (a) specifies the two unit vectors, $\hat{\boldsymbol\epsilon}_1$ and $\hat{\boldsymbol\epsilon}_2$, needed to describe the respective orientations of two dendrimers represented by axisymmetric ellipsoids, $\theta$ being the angle between $\hat{\boldsymbol\epsilon}_1$ and $\hat{\boldsymbol\epsilon}_2$ such that $\cos\theta=\hat{\boldsymbol\epsilon}_1\cdot\hat{\boldsymbol\epsilon}_2$ and $\hat{\boldsymbol{r}}$ being the unit center-to-center vector. Six characteristic pair configurations are listed in panel (b) along with the corresponding values of $\alpha$ and $S$ [see Eqs.~(\ref{eq:alpha}) and (\ref{eq:S})]. Density plots of the conditional distribution function $P(r,\alpha)$ and of the orientational order parameter $S$ are presented in the top and bottom row, respectively; also shown are color codes. The left column of panel (c) shows data for the isolated pair whereas the middle and the right columns
  represent the liquid state at packing fractions $\phi_m=0.199$ and $\phi_m=0.248$, respectively~\cite{remark3}. Black contours are isolines of the pair distribution function $g(r;\alpha)$ at a value slightly smaller than the height of the nearest-neighbor peak.}
\label{gSP} 
\end{center}
\end{figure*}

We now focus on the anisotropic positional and orientational order at close separations. Since acylindricity $c$ is small in all cases explored here, we only monitor the orientation of the dendrimers' long axes relative to the center-to-center vector (Fig.~\ref{gSP}a). Despite this reduction of parameters, the various configurations of two dendrimers can still be described by three angles and therefore a detailed representation of the pair distribution is rather impractical. Instead, we choose to characterize each configuration of two dendrimers using a single quantity
\begin{equation}
\alpha=\frac{1}{2}\left[\left(\hat{\boldsymbol\epsilon}_1\cdot{\hat{\bf r}}\right)^2+\left(\hat{\boldsymbol\epsilon}_2\cdot{\hat{\bf r}}\right)^2\right],
\label{eq:alpha}
\end{equation}
where $\hat{\boldsymbol\epsilon}_1$ and $\hat{\boldsymbol\epsilon}_2$ are the directional unit vectors of the long axes of the dendrimers and $\hat{\bf r}$ is the unit center-to-center vector (Fig.~\ref{gSP}a). Note that $\alpha$ is symmetric with respect to an interchange of dendrimers ($1\leftrightarrow2$) as well as to replacing $\hat{\boldsymbol\epsilon}_i$ by $-\hat{\boldsymbol\epsilon}_i$, thereby reflecting the headless nature of dendrimers. The mapping of the relative orientation of $\hat{\boldsymbol\epsilon}_1, \hat{\boldsymbol\epsilon}_2,$ and ${\hat{\bf r}}$ onto $\alpha$ is not unique as illustrated by the six characteristic configurations shown in the table in Fig.~\ref{gSP}b~\cite{remark2} along with their respective values of $\alpha$. Nonetheless, we find this representation helpful -- much like the radial distribution function can be used to represent the structure of crystals although they are not isotropic.

Additional insight into the relative arrangement of dendrimers is provided by the orientational order parameter 
\begin{equation}
S=\frac{1}{2}\langle 3\cos^2\theta-1\rangle,
\label{eq:S} 
\end{equation}
where $\theta$ is the angle between the long axes (Fig.~\ref{gSP}a) and angular brackets denote an ensemble average. Since the relative orientation of two dendrimers depends on $r$, so does $S$, and as illustrated by the table in Fig.~\ref{gSP}b it distinguishes between some configurations with the same $\alpha$ (e.g., \II \ and \XX). In a pair of perfectly parallel or antiparallel dendrimers, $S=1$ whereas in dendrimers oriented perpendicular to each other $S=-0.5$ (Fig.~\ref{gSP}b).

% INTERACTION OF A DENDRIMER PAIR

In the following we discuss the conditional distribution function $P(r,\alpha)$ and the orientational order parameter $S$ for an isolated pair of dendrimers and a pair of dendrimers in a bulk liquid. These distributions have been normalized for each value of $r$ separately such that the integral over the probabilities across the entire range of $\alpha$ at fixed $r$ is unity. As for $\alpha$, the distributions have been normalized via a random distribution of this variable, generated in simulations. The relative orientation of an isolated pair depends quite dramatically on their separation (left column of Fig.~\ref{gSP}c). The flat profile of $P(r,\alpha)$ for $r/R_g^0\gtrsim3$ (not shown) indicates that at large separations the orientations of the dendrimers are completely uncorrelated such that all orientations of $\hat{\boldsymbol\epsilon}_1$ and $\hat{\boldsymbol\epsilon}_2,$ are equally probable at any $r$ that is large enough. However, as dendrimers penetrate into each other the correlations becomes more and more pronounced: As $r/R_g^0$ drops below $\approx2.5$, $P(r,\alpha)$ peaks at $\
 alpha = 0$ and at $r/R_g^0\approx1.5$ the probability for configurations with $\alpha\gtrsim0.25$ is essentially negligible. In the regime of partial penetration for $1.5\lesssim r/R_g^0\lesssim2.5$, the range of $\alpha$ where $P(r,\alpha)$ is enhanced coincides with the region of positive orientational order parameter (bottom-left panel of Fig.~\ref{gSP}c), indicating the presence of \II \ configurations. The regime of complete overlap for $r/R_g^0\lesssim1$ is characterized by a somewhat broader distribution of $P(r,\alpha)$ peaking at $\alpha\approx0.4$ and excluding the occurrence of states with large $\alpha$ ($\gtrsim0.6)$. Together with the strongly negative orientational order parameter $S$ at small $r$, this implies that overlapping dendrimers show a strong preference to form \TT \ and \LL \ configurations (which are indistinguishable from the \XX \ configuration as $r\to0$). We conclude that repulsion between the overlapping dendrimers forces them into a perpendicular arrangement. 

% DENDRIMER LIQUID

\section{Structure of dendrimer liquid}

In a bulk liquid, the relative orientation of a pair is modified by the local structure. The middle column of Fig.~\ref{gSP}c shows the conditional distribution function $P(r,\alpha)$ (top panel) and the orientational order parameter $S$ at a packing fraction $\phi_m=0.199$~\cite{remark3}; both are strikingly different from their counterparts in two isolated dendrimers. In total, the variations of $P(r,\alpha)$ are less pronounced than in the case of an isolated pair. We note that (i) at the onset of dendrimer-dendrimer interaction at $r/R_g^0\approx2.5$, there is a slight preference for large-$\alpha$ configurations (e.g., \PP); (ii) at intermediate distances $r/R_g^0\approx1.5$ the ($\alpha<0.25$)-configurations are favored and those with large $\alpha$ are increasingly more disfavored just like for the isolated pair; and (iii) overlapping dendrimers ($r\to0$) prefer configurations with $\alpha$ close to 1, e.g. \PP. 

The differences of the relative orientation of a pair of dendrimers in isolation and in a bulk liquid at both small and large separations [(i) and (iii)] are even more pronounced at the larger packing fraction $\phi_m=0.248$ (right column of Fig.~\ref{gSP}c) where particle overlap is rather substantial. This can be readily seen from the pair distribution function $g(r;\alpha)$ represented by an isoline which is superposed onto the plots of $P(r,\alpha)$ and $S$; the value of $g(r;\alpha)$ on the isoline is slightly smaller than the height of the nearest-neighbor peak. This representation is more transparent than a comprehensive set of isolines and  more robust than displaying only the exact location of the nearest-neighbor peak. At any given $\alpha$, the position of the peak is located roughly halfway between by the respective small-$r$ and large-$r$ points on the isoline.     

At the smaller packing fraction $\phi_m=0.199$ (middle column of Fig.~\ref{gSP}c), the tilted $g(r;\alpha)$ isoline shows that the distance between the nearest $\alpha=0$ neighbors is about $1.7R_g^0$; this corresponds to the \II \ configuration since the negative $S=-0.5$ of the \XX \ configuration is inconsistent with the observed positive value of $S$ (bottom-center panel in Fig.~\ref{gSP}c). In contrast, the distance between neighbors in the end-to-end \PP \ configuration with $\alpha=1$ is about $2.25R_g^0$. The intermediate-$\alpha$ configurations (\Idot, \TT, and \LL) are located at a distance of about $2R_g^0$. The relative orientation of two interacting dendrimers in a liquid at the larger packing fraction $\phi_m=0.248$ is qualitatively similar except for the finite probability of dendrimer overlap witnessed by the presence of isolines at small $r$. Since for small $r$ the conditional distribution function $P(r,\alpha)$ peaks at $\alpha=0$, we conclude that in a bulk liquid
  overlapping dendrimers are preferentially parallel to each other.

The portrait of the local structure of a dendrimer liquid presented in Fig.~\ref{gSP}c is markedly different from that characteristic of particles that form liquid crystals. Just like in the nematic phase, the isotropic phase of a liquid-crystalline material consists of rodlike particles locally parallel to each other and the overall isotropic nature of the phase is due to the finite size of these "swarms" of particles and the corresponding finite correlation length. In this phase, (i) the side-to-side \II \ configurations are characterized by a pronounced degree of orientational order, i.e., a large value of $S$ and (ii) the perpendicular intermediate-$\alpha$ configurations (\Idot, \TT, and \LL) should be absent. Thus the contour of the nearest-neighbor shell in the $(r,\alpha)$-plane typical for local nematic order consists of two islands (one at $\alpha\ll 1$ and the other at $\alpha\approx1$) rather than of the diagonal stripe seen both the intermediate- and the large-density ca
 ses presented in Fig.~\ref{gSP}c ($\phi_m=0.199$ and $\phi_m=0.248$), and the degree of order for the nearest-neighbor small-$\alpha$ configurations should be considerably larger than 0. The positive but small value of $S$ of the nearest-neighbor small-$\alpha$ configurations clearly departs from this picture and so does the presence of the intermediate-$\alpha$ configurations in the nearest-neighbor shell. 

% PACKING PATTERN

\section{Packing pattern}

\begin{figure}[t]
\begin{center}
\vspace*{-3mm} 
\includegraphics[width=0.88\columnwidth]{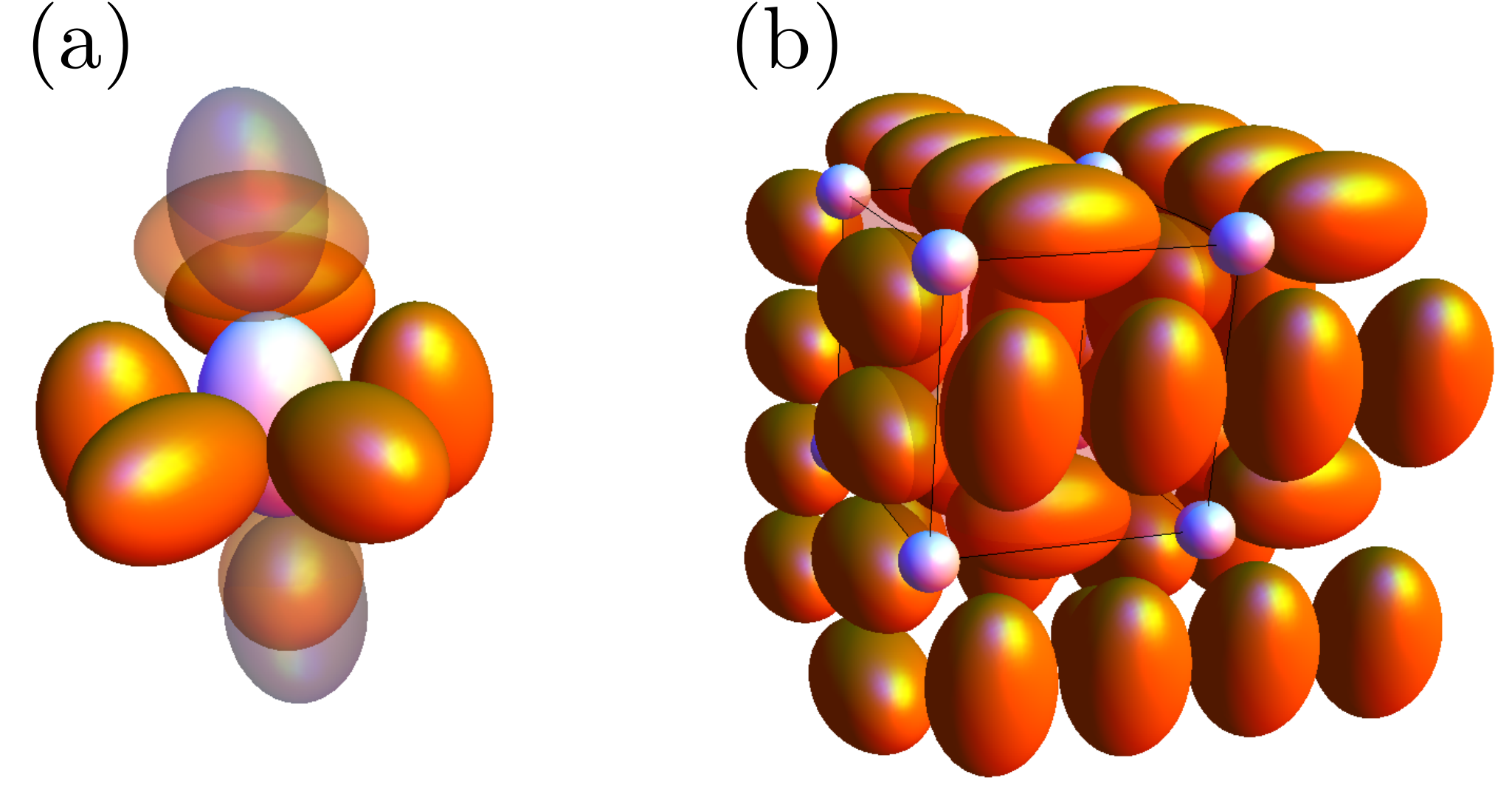}
\end{center}
\caption{Schematic view of the antinematic packing pattern observed in a dendrimer liquid~(a). The nearest-neighbor shell around the reference particle (white ellipsoid) contains \protect\II \ (left and right), \protect\Idot (back), and \protect\TT\ (front) configurations in the equatorial region. The polar regions may be occupied either by \protect\TT\ or \protect\PP\ configurations (gray semitransparent ellipsoids), the latter being a little farther from the reference particle. Panel~(b): The A15 lattice decorated by ellipsoids arranged in three sets of mutually interlocking columns captures the antinematic nature of the pattern shown in panel~(a); the dodecahedral interstitial sites (small white spheres only shown in the unit cell for clarity) lack a preferred orientation.}
\label{packing} 
\end{figure}

These differences suggest that although dendrimers are elongated, they do not align with each other; instead they form a rather specific local structure schematically shown in Fig.~\ref{packing}a. With a discrete rather than a continuous set of dendrimer orientations, the schematic is idealized for clarity. The dendrimers are represented by axisymmetric ellipsoids of aspect ratio of 1.49 consistent with the average ratio of semiaxes of an isolated dendrimer (Fig.~\ref{Phibc}). In the equatorial plane, the central reference dendrimer is surrounded by \II, \Idot, and \TT\ neighbors, the \II \ neighbors being a little closer to the reference particle than the \Idot \ and the \TT \ neighbors. The polar regions are populated either by \TT\ neighbors or by \PP\ neighbors, the latter being somewhat farther from the reference particle than the former. The resulting pattern is thus {\it antinematic}~\cite{Sokalski84} rather than nematic. 

Unlike the origin of nematic order which is usually associated primarily with the particle shape~\cite{Onsager49}, the possible microscopic mechanisms of antinematic order are less clear. Our dendrimers probably favor it because of the combination of elongated shape and softness. If they were spherical, they should form a simple liquid of both single dendrimers and dendrimer clusters as well as a multiple-occupancy crystals~\cite{Mladek06}, which implies that softness alone is not sufficient. In turn, the solid part of the phase diagram of hard ellipsoids is dominated by an unusual simple monoclinic packing~\cite{Radu09} somewhat reminiscent of the combination of \II \ and \XX \ configurations present in the nearest-neighbor shell -- but only if the aspect ratios larger than about 2, i.e., in particles that are more elongated than our dendrimers. This means that the antinematic order is not induced by the shape alone either. We are led to conjecture that both softness and elongation 
 of the particles are required for an antinematic local structure.

% MANY-BODY EFFECTS

The distinct differences between $P(r,\alpha)$ for an isolated pair of dendrimers and two interacting macromolecules in a bulk liquid (top row of Fig.~\ref{gSP}c) are a clear signature of many-body effects. Equally telling is the comparison of the panels in the bottom row of Fig.~\ref{gSP}c showing the orientational order of the dendrimers. In the case of an isolated pair the overlapping dendrimers are arranged perpendicular to each other (\XX) as witnessed by the extended red region in the bottom-left panel in Fig.~\ref{gSP}c. In contrast, in a bulk liquid they are arranged end-to-end (\PP) as argued above. Such a strong effect would not be possible unless the local "cage" of neighbors were tight and ordered enough, leaving little space for a perpendicular orientation of overlapping dendrimers.

Our results are qualitatively consistent with the nature of orientational order in crystals of hard ellipsoids~\cite{Pfleiderer07} and deformable hard spheres~\cite{Batista10}. At large enough semiaxes ratio, the former were found to form a simple monoclinic phase with 2 ellipsoids per unit cell (SM2) such that the angle between their long axes is nonzero~\cite{Pfleiderer07}, showing that anisometric particles may prefer nontrivial local orientational arrangement even in case of simple hard interaction. On the other hand, at large densities the minimal-energy structure of deformable hard spheres is a layered crystal referred to as S2 and consisting of ellipsoidal particles, their orientation alternating by $90^\circ$ from layer to layer~\cite{Batista10}. Here each particle has a total of 14 nearest neighbors: 2 of them are of \II \ type, 8 of \TT \ type, and 4 of \DD \ type with $\alpha = 0.5$ and $S=1$; note that the \DD \ neighbors are absent in our dendrimer liquid. Thus 
 both hard
  ellipsoids and deformable hard spheres crystallize in lattices characterized by nontrivial orientational order, and from the comparison of SM2 and S2 lattices it appears that the degree of misalignment of nearest neighbors in deformable particles is larger than in undeformable particles. However, even the more misaligned S2 lattice still contains a large fraction of parallel nearest neighbors compared to our dendrimer liquid. In this sense, our antinematic local order differs qualitatively from the previously reported model colloidal structures. Moreover, the numerical framework used here is more detailed and realistic than those in Refs.~\cite{Pfleiderer07} and \cite{Batista10} and the model dendrimers studied are intrinsically anisometric and deformable as well as interpenetrable at the same time. In view of the results reported in Refs.~\cite{Pfleiderer07} and \cite{Batista10}, we infer that the antinematic order is induced by a combination of these particle properties.

\section{Conclusions}

Based on these observations we can formulate our expectations for the structure of dendrimer crystals at even higher packing fractions. As the average orientation of dendrimers with a local antinematic order is fairly isotropic, a crystalline lattice formed by dendrimers is likely highly symmetric, i.e., cubic. However, the single-site-type cubic lattices (e.g., simple, body-centered, and face-centered) are incompatible with antinematic order, implying that dendrimer crystals must be more complex. An obvious candidate is the A15 lattice~\cite{Balagurusamy97,Zeng04} based on three sets of mutually perpendicular columns of particles which accommodate many features of the packing pattern (Fig.~\ref{packing}b) discussed above. Each columnar site (ellipsoids in Fig.~\ref{packing}b) has \II \ neighbors and hybrids of \Idot, \TT, and \LL \ neighbors as well as the more distant \PP \ neighbors. Unlike any single-site cubic crystal structure, the A15 lattice is consistent with antinematic order although the interstitial sites (small white spheres in Fig.~\ref{packing}b) are characterized by a dodecahedral environment and are most easily populated by spherical rather than elongated particles. Thus the stability of the A15 lattice may be directly related to the elongated shape and deformability of dendrimers.  

Our investigations provide for the first time unambiguous evidence about the origin of the antinematic order itself. In contrast to hard rodlike particles where excluded-volume interactions induce the {\it nematic} phase~\cite{Bolhuis97}, we identify the softness and the anisometry of the particles as {\it the} key mechanisms that are responsible for {\it antinematic} local particle arrangements. The dramatic change of the behavior due to softness is very reminiscent of the characteristic differences between the phase diagrams of hard and soft spheres~\cite{Mladek06,Prestipino07,Pamies09}. In this context, a very interesting question raised by our results is the possible existence of an antinematic phase with long-range order. The search for such a structure will entail a detailed examination of many types of deformable particles and has been relegated to future work.

\section{Acknowledgments}
We are indebted to Ronald Blaak (Universit\"at Wien), Hiroshi Noguchi (University of Tokyo), Silvano Romano (Universit\` a di Pavia), and Gregor Ska\v cej (University of Ljubljana) for many helpful suggestions. This work was supported by the Marie-Curie Initial Training Network COMPLOIDS (FP7-PEOPLE-ITN-2008 Grant No.~234810), by the Slovenian Research Agency (Grant No.~P1-0055), and by the Austrian Science Fund (FWF) under Projects Nos. P23910-N16 and F41 (SFB ViCoM).


\begin{thebibliography}{99}
\bibitem{remark2} The orientation of the center-to-center unit vector is given implicitly in all of these configurations except \Idot \ where we assume that the angle between the vertical dendrimer and the center-to-center vector is $45^\circ$ like in the \LL \ configuration and in the \DD \ configuration discussed below.
\end{thebibliography}
\end{document}